# Quantum states with less energy than the vacuum in Dirac Hole Theory

by

Dan Solomon

Rauland-Borg Corporation
3450 W Oakton
Skokie, IL 60076
USA

Email: dan.solomon@rauland.com

March 20, 2007

**Abstract**

In Dirac's hole theory the vacuum state is generally believed to be the state of minimum energy. However it has recently been shown that this is not the case. In [6] it was shown that energy can be extracted from the hole theory vacuum state through the application of an electric field so that the final state has less energy than that of the vacuum state. In this paper we will confirm the results of [6] by calculating the change in the energy of the vacuum state due to its interaction with a specific electric field. It will be shown that the final state has less energy than the original vacuum state.

PACS Nos: 03.65-w, 11.10-z

# 1. Introduction.

It is well known that there are both positive and negative energy solutions to the Dirac equation. This creates a problem in that an electron in a positive energy state will quickly decay into a negative energy state in the presence of perturbations. This, of course, is not normally observed to occur. This problem is resolved in Dirac's hole theory (HT) by assuming that all the negative energy states are occupied by a single electron and then evoking the Pauli exclusion principle to prevent the decay of the positive energy electrons into negative energy states.

In HT the vacuum state is the state in which each unperturbed negative energy solution to the Dirac equation is assumed to be occupied by a single electron and each positive energy state is unoccupied. The electrons in these unperturbed negative energy states, the so called *Dirac sea*, are assumed to be unobservable. What we observe are variations from the unperturbed vacuum state.

Now consider a "simple" quantum theory consisting of non-interacting electrons in a background classical electromagnetic field. In this case it is generally assumed that quantum field theory (QFT) and hole theory (HT) will give identical results. However, recently, several papers have appeared in the literature pointing out that there are differences between HT and QFT for this simple situation (see [1][2][3][4][5][6]). One somewhat surprising result of this research was derived in [6] where it was shown that the HT vacuum state is not the state of minimum energy. That is, in HT, there exist states with less energy than that of the vacuum state. This result was unexpected because it is generally assumed that the HT vacuum state is the state of minimum energy. This result is also in sharp contrast to the standard formulation of QFT where the vacuum state is the minimum energy state.

In this article we will confirm the results derived in [6] using a different approach. In [6] the change in the energy due to the interaction of the initial vacuum state with a specific electric field was calculated. It was shown that the final state had less energy then the initial vacuum state therefore there exist states with less energy then the vacuum state in HT. In [6] perturbation theory was used to calculate the change in the wave functions of the initial electrons. The reason for this was that it was not possible to find an exact solution to Dirac's equation for the electric field specified in [6]. In this article a different approach will be used. We will determine the change in the energy of the initial vacuum state due to an electric field for which an exact solution to Dirac's equation can be determined. It will be shown that the final state has less energy than the initial vacuum state. This result is consistent with [6] and shows that there must exist states with less energy than the vacuum state in HT. The value of the alternative approach used in this paper is that it shows the results of [6] are not due to some artifact of the perturbation method which was used in [6] to derive the results.

## 2. The Dirac Equation.

In order to simplify the discussion and avoid unnecessary mathematical details we will assume that the electrons are non-interacting, i.e., they only interact with an external electric potential. Also we will work in 1-1 dimensional space-time where the space dimension is taken along the z-axis and use natural units so that $\hbar = c = 1$. In this case the Dirac equation for a single electron in the presence of an external electric potential is,

$$i\frac{\partial \psi(z,t)}{\partial t} = H\psi(z,t) \quad (2.1)$$

where the Dirac Hamiltonian is given by,

$$H = H_0 + qV(z,t) \quad (2.2)$$

where $H_0$ is the Hamiltonian in the absence of interactions, $V(z,t)$ is an external electrical potential, and q is the electric charge. For the 1-1D case,

$$H_0 = \left(-i\sigma_x \frac{\partial}{\partial z} + m\sigma_z\right) \quad (2.3)$$

where $\sigma_x$ and $\sigma_z$ are the usual Pauli matrices. In this case the orthonormal free field solutions (V is zero) of (2.1) are given by,

$$\varphi^{(0)}_{\lambda,p_r}(z,t) = \varphi^{(0)}_{\lambda,p_r}(z)\exp\left(-i\varepsilon^{(0)}_{\lambda,p_r}t\right) = u_{\lambda,p_r}\exp\left(-i\left(\varepsilon^{(0)}_{\lambda,p_r}t - p_r z\right)\right) \quad (2.4)$$

where $p_r = 2\pi r/L$ is the momentum, $L$ is the 1-dimensional integration volume which is assumed to approach infinity, $r$ is an integer, $\lambda = \pm 1$ is the sign of the energy and where,

$$\varepsilon^{(0)}_{\lambda,p_r} = \lambda E \; ; \; E = +\sqrt{p_r^2 + m^2} \; ; \; u_{\lambda,p_r} = \frac{N_{\lambda,p_r}}{\sqrt{L}}\begin{pmatrix} 1 \\ p_r/(\lambda E + m) \end{pmatrix} \; ; \; N_{\lambda,p_r} = \sqrt{\frac{\lambda E + m}{2\lambda E}} \quad (2.5)$$

The quantities $\varphi^{(0)}_{\lambda,p_r}(z)$ satisfy the relationship,

$$H_0 \varphi^{(0)}_{p_r,r}(z) = \varepsilon^{(0)}_{\lambda,p_r} \varphi^{(0)}_{\lambda,p_r}(z) \qquad (2.6)$$

The $\varphi^{(0)}_{\lambda,r}(z)$ form an orthonormal basis set and satisfy,

$$\int_{-L/2}^{+L/2} \varphi^{(0)\dagger}_{\lambda,p_r}(z) \varphi^{(0)}_{\lambda',p_s}(z) dz = \delta_{\lambda\lambda'} \delta_{p_r,p_s} \qquad (2.7)$$

The energy of a normalized wave function $\varphi(z,t)$ is defined by,

$$\varepsilon(\psi(z,t)) = \int_{-L/2}^{+L/2} \varphi^\dagger(z,t)(H_0 + V(z,t)) \varphi(z,t) dz \qquad (2.8)$$

### **4. Time varying perturbation**

In this section we will examine the effect of a time varying electric potential on the hole theory vacuum state. Assume that, at some initial time $t_0$, the electric potential is zero and that the system is in the unperturbed vacuum state. This is the state where each negative energy wave function $\varphi^{(0)}_{-1,p}$ is occupied by a single electron and each positive energy state $\varphi^{(0)}_{+1,p}$ is unoccupied. Now, consider the change in the energy due to an interaction with an external electric potential which is applied at some time $t > t_0$ and then removed at some later time $t_1$ so that,

$$V(z,t) = 0 \text{ for } t \leq t_0; \quad V(z,t) \neq 0 \text{ for } t_0 < t < t_1; \quad V(z,t) = 0 \text{ for } t \geq t_1 \qquad (3.1)$$

Now what is the change in the energy of the quantum system due to this interaction? Under the action of the electric potential each initial wave function $\varphi^{(0)}_{\lambda,p}(z,t_0)$ evolves according to the Dirac equation into the final state $\varphi_{\lambda,p}(z,t_1)$. The change in the energy of the state $\varphi_{\lambda,p}(z,t)$ from $t_0$ to $t_1$ is given by,

$$\delta\varepsilon_{\lambda,p}(t_0 \to t_1) = \varepsilon_{\lambda,p}(t_1) - \varepsilon_{\lambda,p}(t_0) \qquad (3.2)$$

Use (2.8) along with the $V(z,t_1) = V(z,t_0) = 0$ (from (3.1)) in the above to obtain,

$$\delta\varepsilon_{\lambda,p}(t_0 \to t_1) = \int_{-L/2}^{+L/2} \varphi^\dagger_{\lambda,p}(z,t_1) H_0 \varphi_{\lambda,p}(z,t_1) dz - \int_{-L/2}^{+L/2} \varphi^\dagger_{\lambda,p}(z,t_0) H_0 \varphi_{\lambda,p}(z,t_0) dz \qquad (3.3)$$

Note that $\varphi_{\lambda,p}(z,t_0) = \varphi_{\lambda,p}^{(0)}(z,t_0)$. Use this along with (2.6) and (2.7) in the above to obtain,

$$\delta\varepsilon_{\lambda,p}(t_0 \to t_1) = \int_{-L/2}^{+L/2} \varphi_{\lambda,p}^{\dagger}(z,t_1) H_0 \varphi_{\lambda,p}(z,t_1) dz - \lambda E \qquad (3.4)$$

As discussed in [6] the change in the energy of the vacuum state from $t_0$ to $t_1$ is the sum of change in energy of each vacuum electron and is therefore given by,

$$\Delta\xi_{vac}(t_0 \to t_1) = \sum_r \delta\varepsilon_{-1,p_r}(t_0 \to t_1) \qquad (3.5)$$

In the limit $L \to \infty$ we can make the substitution $\sum_r \to \int_{-\infty}^{+\infty} \frac{Ldp}{2\pi}$ to obtain,

$$\Delta\xi_{vac}(t_0 \to t_1) = \int_{-\infty}^{+\infty} \frac{Ldp}{2\pi} \delta\varepsilon_{-1,p}(t_0 \to t_1) = \int_0^{+\infty} \frac{Ldp}{2\pi} \left(\delta\varepsilon_{-1,p}(t_0 \to t_1) + \delta\varepsilon_{-1,-p}(t_0 \to t_1)\right) \qquad (3.6)$$

In order to calculate $\Delta\xi_{vac}(t_0 \to t_1)$ it is necessary to calculate $\delta\varepsilon_{-1,p}(t_0 \to t_1)$ for all $p$.

We shall determine $\delta\varepsilon_{-1,p}$ for the electric potential given by,

$$V(z,t) = -zc\alpha(1-\theta(t))e^{-ct} \qquad (3.7)$$

where $\alpha$ and $c$ are constants with $c < 0$ and $\alpha < 0$ and,

$\theta(t)$ is 1 for $t \geq 0$ and 0 for $t < 0$ \qquad (3.8)

The electric field corresponding to this potential is given by,

$$-\partial V(z,t)/\partial z = c\alpha(1-\theta(t))e^{-ct} \qquad (3.9)$$

The potential is zero at the initial time $t_0 = -\infty$ (because $e^{-ct_0} \to 0$ due to the fact that $c < 0$) and removed at the final time $t_1 = 0$ (because $(1-\theta(t)) = 0$ for $t \geq 0$). We want to determine the change in the energy from initial to final time of each vacuum electron.

The solution to the Dirac equation for the potential (3.7) for electrons that are initially in the unpertubed state $\varphi_{\lambda,p}^{(0)}(z,t_0)$ is shown in Appendix 1 to be, for $t < 0$,

$$\varphi_{\lambda,p}(z,t) = \frac{\eta_{\lambda,p}}{\sqrt{L}} \begin{pmatrix} C_{\lambda,p}(t) \\ D_{\lambda,p}(t) \end{pmatrix} e^{i(p-A(t))z} \qquad (3.10)$$

and, for $t \geq 0$,

$$\varphi_{\lambda,p}(z,t) = e^{-iH_0 t}\varphi_{\lambda,p}(z,0) \qquad (3.11)$$

where $\varphi_{\lambda,p}(z,0)$ in the (3.11) equals $\varphi_{\lambda,p}(z,t)$ in (3.10) with $t=0$. This ensures that the wave function is continuous at $t=0$. The various quantities in the above equations are defined by,

$$C_{\lambda,p}(t) = e^{-iR/2} R^{i\lambda\mu} \begin{bmatrix} (m-p+\lambda E)\phi(J,K,iR) \\ +cR\phi'(J,K,iR) \end{bmatrix} \tag{3.12}$$

$$D_{\lambda,p}(t) = e^{-iR/2} R^{i\lambda\mu} \begin{bmatrix} (m+p-\lambda E)\phi(J,K,iR) \\ -cR\phi'(J,K,iR) \end{bmatrix} \tag{3.13}$$

where,

$$J = \frac{i}{c}(\lambda E - p); \quad K = 1 + \frac{2i\lambda E}{c}; \quad \mu = \frac{E}{c}; \quad R(t) = \frac{2\alpha}{c} e^{-ct}; \quad A(t) = \alpha e^{-ct} = \frac{cR(t)}{2} \tag{3.14}$$

and,

$$\eta_{\lambda,p} = e^{-i\lambda\mu Log(2\alpha/c)} \sqrt{\frac{1}{4E(E-\lambda p)}} \tag{3.15}$$

and where $\phi(J,K,iR)$ is the confluent hypergeometric function which is given by,

$$\phi(J,K,iR) = 1 + \frac{J}{L}iR + \frac{J(J+1)}{K(K+1)} \frac{(iR)^2}{2!} + \ldots \tag{3.16}$$

and $\phi'(J,K,iR)$ is the first derivative of $\phi(J,K,iR)$ with respect to $iR$. In deriving the above result we have used results developed in chapter 3 of [7].

Now, as shown in Appendix 1, at the initial time $t_0 = -\infty$ the state $\varphi_{\lambda,p}(z,t_0)$ is equal to the initial unperturbed state $\varphi_{\lambda,p}^{(0)}(z,t_0)$. Since the electric potential approaches zero at $t_0$ we have that the energy of the state at $t_0$ is $\varepsilon_{\lambda,p}(t_0) = \lambda E$. This, of course, is the energy of the initial unperturbed state. Under the action of the electric potential the initial state $\varphi_{\lambda,p}^{(0)}(z,t_0)$ evolves into the final state $\varphi_{\lambda,p}(z,t_1)$ where $t_1 = 0$ and where $\varphi_{\lambda,p}(z,t_1)$ is given by the expression (3.10) with $t=0$. Use $\varphi_{\lambda,p}(z,0)$ from (3.10) along with the fact that $V(z,0) = 0$ in (2.8) to obtain the following expression for the energy of the state $\varphi_{\lambda,p}(z,t_1)$,

$$\varepsilon_{\lambda,p}(t_1) = |\eta_{\lambda,p}|^2 \begin{bmatrix} (p - A(t_1))(C(t_1)^* D(t_1) + C(t_1) D(t_1)^*) \\ + m(C(t_1)^* C(t_1) - D(t_1)^* D(t_1)) \end{bmatrix} \quad (3.17)$$

Use (3.12) and (3.13) to obtain,

$$C(t_1)^* D(t_1) + C(t_1) D(t_1)^* = 2(m^2 - (p - \lambda E)^2) \phi^* \phi \\ + 2cR(p - \lambda E)(\phi^* \phi' + \phi'^* \phi) - 2c^2 R^2 \phi'^* \phi' \quad (3.18)$$

and

$$C(t_1)^* C(t_1) - D(t_1)^* D(t_1) = -4m(p - \lambda E)\phi^* \phi + 2cR(t_1) m(\phi'^* \phi + \phi^* \phi') \quad (3.19)$$

where we write $\phi$ instead of $\phi(J, K, iR(t_1))$. Use these results in (3.17) to obtain,

$$\varepsilon_{\lambda,p}(t_1) = |\eta_{\lambda,p}|^2 \left\{ \begin{array}{l} (p - A(t_1)) \begin{pmatrix} 2(m^2 - (p - \lambda E)^2) \phi^* \phi \\ +2cR(t_1)(p - \lambda E)(\phi^* \phi' + \phi'^* \phi) - 2c^2 R(t_1)^2 \phi'^* \phi' \end{pmatrix} \\ -4m^2(p - \lambda E)\phi^* \phi + 2cR(t_1) m^2 (\phi^* \phi' + \phi'^* \phi) \end{array} \right\} \quad (3.20)$$

Use (3.14) and rearrange terms to obtain,

$$\varepsilon_{\lambda,p}(t_1) = |\eta_{\lambda,p}|^2 \left\{ \begin{array}{l} \phi^* \phi \left[ 2(m^2 - (p - \lambda E)^2)\left(p - \frac{c}{2}R(t_1)\right) - 4m^2(p - \lambda E) \right] \\ +2cR(t_1)\left( m^2 + (p - \lambda E)\left(p - \frac{c}{2}R(t_1)\right) \right)(\phi^* \phi' + \phi'^* \phi) \\ -2c^2 R(t_1)^2 \left(p - \frac{c}{2}R(t_1)\right) \phi'^* \phi' \end{array} \right\} \quad (3.21)$$

Rearrange terms further to obtain,

$$\varepsilon_{\lambda,p}(t_1) = |\eta_{\lambda,p}|^2 \left\{ \begin{array}{l} 4\phi^* \phi \lambda (E - \lambda p)\left( E^2 - \frac{cpR(t_1)}{2} \right) \\ +2cR(t_1)(E - \lambda p)\left( E + \frac{c\lambda R(t_1)}{2} \right)(\phi^* \phi' + \phi'^* \phi) \\ -2c^2 R(t_1)^2 \left( p - \frac{cR(t_1)}{2} \right) \phi'^* \phi' \end{array} \right\} \quad (3.22)$$

The change in the energy from the initial time $t_0 = -\infty$ to the final time $t_1 = 0$ is $\delta \varepsilon_{\lambda,p}(t_0 \to t_1) = \varepsilon_{\lambda,p}(0) - \varepsilon_{\lambda,p}(-\infty)$. Based on the form of the confluent hypergeomtric function it is evident that we can express $\varepsilon_{\lambda,p}(t)$ as a series expansion in powers of the parameter $\alpha$. Therefore we will evaluate $\delta\varepsilon_{\lambda,p}$ in the limit $\alpha \to 0$ and keep the lowest order terms. Since $R(t)$ is proportional to $\alpha$ this is equivalent to an expansion in powers of $R(t)$.

Use (3.16) in (3.21) and after some algebraic manipulation and the help of the expressions in Appendix 2 we obtain,

$$\varepsilon_{\lambda,p}(t_1) = \varepsilon_{\lambda,p}^{(0)}(t_1) + \varepsilon_{\lambda,p}^{(1)}(t_1) + \varepsilon_{\lambda,p}^{(2)}(t_1) + O(\alpha^3) \tag{3.23}$$

where $O(\alpha^3)$ means terms to the third power of $\alpha$ or higher and where,

$$\varepsilon_{\lambda,p}^{(0)}(t_1) = 4\lambda |\eta_{\lambda,p}|^2 (E - \lambda p) E^2 = \lambda E \tag{3.24}$$

$$\varepsilon_{\lambda,p}^{(1)}(t_1) = |\eta_{\lambda,p}|^2 R(t_1) \begin{bmatrix} 4\lambda(E - \lambda p)\left(E^2 i \left(\frac{J}{K} - c.c\right) - \frac{cp}{2}\right) \\ +2cE(E - \lambda p)\left(\left(\frac{J}{K} + c.c\right)\right) \end{bmatrix} \tag{3.25}$$

$$\varepsilon_{\lambda,p}^{(2)}(t_1) = \varepsilon_{a;\lambda,p}^{(2)}(t_1) + \varepsilon_{b;\lambda,p}^{(2)}(t_1) + \varepsilon_{c;\lambda,p}^{(2)}(t_1) \tag{3.26}$$

where,

$$\varepsilon_{a;\lambda,p}^{(2)}(t_1) = 4\lambda |\eta_{\lambda,p}|^2 R(t_1)^2 (E - \lambda p) \begin{bmatrix} E^2 \left(\left|\frac{J}{K}\right|^2 - \left(\frac{1}{2}\right)\left(\frac{J(J+1)}{K(K+1)} + c.c\right)\right) \\ -i\frac{cp}{2}\left(\frac{J}{K} - c.c\right) \end{bmatrix} \tag{3.27}$$

$$\varepsilon_{b;\lambda,p}^{(2)}(t_1) = 2c |\eta_{\lambda,p}|^2 R(t_1)^2 (E - \lambda p) \left[ iE\left(\frac{J(J+1)}{K(K+1)} - c.c\right) + i\frac{c\lambda}{2}\left(\frac{J}{K} + c.c\right) \right] \tag{3.28}$$

and,

$$\varepsilon_{c;\lambda,p}^{(2)}(t_1) = -2c^2 p |\eta_{\lambda,p}|^2 R(t_1)^2 \left|\frac{J}{K}\right|^2 \tag{3.29}$$

where c.c. is the complex conjugate of the previous term. Note that $\varepsilon_{\lambda,p}^{(0)}(t_1)$ is just the energy of the initial unperturbed state.

Use the above expressions along with the relationships in Appendix 3 to obtain,

$$\varepsilon_{\lambda,p}^{(1)}(t_1) = -2\lambda c p |\eta_{\lambda,p}|^2 (E - \lambda p) R(t_1) = \frac{-\lambda p \alpha}{E} e^{-ict_1} \tag{3.30}$$

$$\varepsilon_{\lambda,p}^{(2)}(t_1) = \frac{\lambda c^2 m^2 R(t_1)^2}{(2E)(4E^2 + c^2)} = \frac{4\lambda m^2 \alpha^2 e^{-2ct_1}}{(2E)(4E^2 + c^2)} \tag{3.31}$$

Use the fact that $t_1 = 0$ to obtain,

$$\varepsilon_{\lambda,p}(t_1 = 0) = \lambda E - \frac{\lambda p \alpha}{E} + \frac{4\lambda m^2 \alpha^2}{(2E)(4E^2 + c^2)} + O(\alpha^3) \tag{3.32}$$

Recall that $\varepsilon_{\lambda,p}(t_0 = -\infty) = \lambda E$. Therefore we obtain that the change in energy due to the application of the electric potential is,

$$\delta\varepsilon_{\lambda,p}(t_0 \to t_1) = \varepsilon_{\lambda,p}(0) - \varepsilon_{\lambda,p}(-\infty) = \frac{-\lambda p \alpha}{E} + \frac{4\lambda m^2 \alpha^2}{(2E)(4E^2 + c^2)} + O(\alpha^3) \tag{3.33}$$

Use the above to obtain,

$$\delta\varepsilon_{-1,p}(t_0 \to t_1) + \delta\varepsilon_{-1,-p}(t_0 \to t_1) = -\frac{4\alpha^2 m^2}{E(4E^2 + c^2)} + O(\alpha^3) \tag{3.34}$$

Use this in (3.6) to obtain,

$$\Delta\xi_{vac}(t_0 \to t_1) = -\alpha^2 \int_0^{+\infty} \frac{L dp}{2\pi} \left( \frac{4m^2}{E(4E^2 + c^2)} \right) + O(\alpha^3) \tag{3.35}$$

In the limit that $\alpha \to 0$ it is evident that we can drop the $O(\alpha^3)$ and $\Delta\xi_{vac}(t_0 \to t_1)$ will be negative due to the fact that the integrand of the above expression is positive. Therefore the energy has been extracted from the initial vacuum state due to application of the electric field. This is consistent with the result of [6].

## 4 Conclusion.

In conclusion, we have examined the vacuum in Dirac's hole theory. The vacuum electrons obey the Dirac equation and the energy of these electrons will change in response to an applied electric field. We have calculated the change in the energy of the vacuum electrons due to the electric potential given by equation (3.7) in the limit $\alpha \to 0$. It has been shown that in this case the total change in the energy of the vacuum electrons is negative. Therefore energy has been extracted from the initial vacuum state due to its interaction with the electric field and the final state has less energy then the initial unperturbed vacuum state. This is consistent with the results of [6] and shows that the HT theory vacuum state is not the state of minimum energy.

## **Appendix 1**

We will show that (3.10) is a solution to the Dirac equation with the electric potential given by (3.7) and the initial condition $\varphi_{\lambda,p}(z,-\infty) = \varphi^{(0)}_{\lambda,p}(z,-\infty)$. For $t<0$ the Dirac equation for the potential of (3.7) is given by,

$$i\frac{\partial \varphi_{\lambda,p}(z,t)}{\partial t} = \left(-i\sigma_x \frac{\partial}{\partial z} + m\sigma_z\right)\varphi_{\lambda,p}(z,t) - zc\alpha e^{-ct}\varphi_{\lambda,p}(z,t) \tag{A.1}$$

Use (3.10) in the above to obtain,

$$i\frac{\partial}{\partial t}\begin{pmatrix} C_{\lambda,p} \\ D_{\lambda,p} \end{pmatrix} = \begin{pmatrix} D_{\lambda,p} \\ C_{\lambda,p} \end{pmatrix}(p - A(t)) + m\begin{pmatrix} C_{\lambda,p} \\ -D_{\lambda,p} \end{pmatrix} \tag{A.2}$$

where $A(t)$ is defined in (3.14). Now it can be shown that,

$$C_{\lambda,p} = (m - p + A(t))\chi_{\lambda,p} + i\frac{\partial \chi_{\lambda,p}}{\partial t} \tag{A.3}$$

and,

$$D_{\lambda,p} = (m + p - A(t))\chi_{\lambda,p} - i\frac{\partial \chi_{\lambda,p}}{\partial t} \tag{A.4}$$

where $\chi_{\lambda,p}$ is a yet to be determined function. When the above relationships are substituted into (A.2) we obtain,

$$\frac{\partial^2 \chi_{\lambda,p}}{\partial t^2} + \left(m^2 + (p - A)^2 - i\frac{\partial A}{\partial t}\right)\chi_{\lambda,p} = 0 \tag{A.5}$$

Use $\partial/\partial t = (\partial R/\partial t)\partial/\partial R$ and $\partial R/\partial t = -cR$ to obtain,

$$c^2 R\frac{\partial}{\partial R}\left(R\frac{\partial \chi_{\lambda,p}}{\partial R}\right) + \left(m^2 + \left(p - \frac{cR}{2}\right)^2 + i\frac{c^2 R}{2}\right)\chi_{\lambda,p} = 0 \tag{A.6}$$

Let,

$$\chi_{\lambda,p} = e^{-iR/2}R^{i\lambda\mu}W_{\lambda,p} \tag{A.7}$$

where $W_{\lambda,p}$ is evaluated in the following. Use this in (A.6) to obtain,

$$(iR)\frac{\partial^2 W_{\lambda,p}}{\partial (iR)^2} + \left[\left(1 + \frac{2i\lambda E}{c}\right) - (iR)\right]\frac{\partial W_{\lambda,p}}{\partial (iR)} - \frac{i(\lambda E - p)}{c}W_{\lambda,p} = 0 \tag{A.8}$$

Use (3.14) in the above to obtain,

$$(iR)\frac{\partial^2 W_{\lambda,p}}{\partial (iR)^2} + \left[K - (iR)\right]\frac{\partial W_{\lambda,p}}{\partial (iR)} - JW_{\lambda,p} = 0 \tag{A.9}$$

The solution to this equation is (See Section 9.2 of [8]),

$$W_{\lambda,p} = \phi(J, K, iR) \tag{A.10}$$

Therefore,

$$\chi_{\lambda,p} = e^{-iR/2} R^{i\lambda\mu} \phi(J, K, iR) \tag{A.11}$$

Use this in (A.3) and (A.4) to show that (3.12) and (3.13) are valid.

Now we have to show that (3.10) obeys the initial condition. The initial condition is that at $t \to -\infty$,

$$\varphi_{\lambda,p}(z, -\infty) = \varphi_{\lambda,p}^{(0)}(z, -\infty) \tag{A.12}$$

At $t \to -\infty$ we have $R(t) \to 0$ so $\phi(J, K, iR) \to 1$. Also note that,

$$R^{i\lambda\mu} = e^{i\lambda\mu Log R} = e^{i\lambda\mu Log(2\alpha/c)} e^{-i\lambda Et} \tag{A.13}$$

We this in (3.12) and (3.13) to obtain,

$$C_{\lambda,p}(t \to -\infty) = e^{i\lambda\mu Log(2\alpha/c)} e^{-i\lambda Et} (m - p + \lambda E) \tag{A.14}$$

$$D_{\lambda,p}(t \to -\infty) = e^{i\lambda\mu Log(2\alpha/c)} e^{-i\lambda Et} (m + p - \lambda E) \tag{A.15}$$

Therefore,

$$\varphi_{\lambda,p}(z, t \to -\infty) = \frac{e^{-i\lambda Et}}{\sqrt{L}} \sqrt{\frac{1}{4E(E - \lambda p)}} \binom{(m - p + \lambda E)}{(m + p - \lambda E)} e^{ipz} \tag{A.16}$$

With some algebraic manipulation his can be shown to be equal to $\varphi_{\lambda,p}^{(0)}(z,t)$. Therefore (3.10) obeys the Dirac equation and satisfies the correct initial conditions. Now at $t \geq 0$ the electric potential is zero so the Dirac equation is,

$$i\frac{\partial \varphi_{\lambda,p}(z,t)}{\partial t} = H_0 \varphi_{\lambda,p}(z,t) \tag{A.17}$$

The solution to this is trivially shown to be,

$$\varphi_{\lambda,p}(z,t) = e^{-iH_0 t} \psi(z) \text{ for } t \geq 0 \tag{A.18}$$

where $\psi(z)$ determined from the fact the solution must be continuous across the boundary condition at $t = 0$. Therefore $\psi(z) = \varphi_{\lambda,p}(z,0)$ where $\varphi_{\lambda,p}(z,0)$ is given by (3.10) with $t = 0$.

## Appendix 2

Using (3.16) and the fact that $R(t)$ is proportional to $\alpha$ we obtain,

$$\phi = 1 + \frac{J}{K}iR + \frac{J(J+1)}{K(K+1)}\frac{(iR)^2}{2!} + O(\alpha^3) \tag{B.1}$$

and

$$\phi' = \frac{J}{K} + \frac{J(J+1)}{K(K+1)}iR + O(\alpha^2) \tag{B.2}$$

Use these results to obtain the following relationships,

$$\phi^*\phi = 1 + \left(\frac{J}{K} - c.c.\right)iR - \left(\frac{J(J+1)}{K(K+1)} + c.c.\right)\frac{(R)^2}{2!} + \left|\frac{J}{K}\right|^2 (R)^2 + O(\alpha^3) \tag{B.3}$$

$$\phi^*\phi' + c.c. = \left(\frac{J}{K} + c.c.\right) + \left(\frac{J(J+1)}{K(K+1)} - c.c.\right)iR + O(\alpha^2) \tag{B.4}$$

$$\phi'^*\phi' = \left|\frac{J}{K}\right|^2 + O(\alpha) \tag{B.5}$$

We can use these equations in (3.22) to obtain (3.24) through (3.27).

## Appendix 3

Use (3.14) to obtain the following relationships,

$$\frac{J}{K} + \frac{J^*}{K^*} = 4\left(\frac{E(E-\lambda p)}{c^2 + 4E^2}\right) \tag{C.1}$$

$$\frac{J}{K} - \frac{J^*}{K^*} = \frac{2\lambda ic(E-\lambda p)}{(c^2 + 4E^2)} \tag{C.2}$$

$$\frac{J(J+1)}{K(K+1)} = \frac{(\lambda E - p)}{2cD(E)}\left(1 + \frac{i(\lambda E - p)}{c}\right)\left(i\left(1 - \frac{2E^2}{c^2}\right) + \frac{3\lambda E}{c}\right) \tag{C.3}$$

where,

$$D(E) = \left(1 + \frac{4E^2}{c^2}\right)\left(1 + \frac{E^2}{c^2}\right) \tag{C.4}$$

From this we obtain,

$$\frac{J(J+1)}{K(K+1)} + c.c. = \frac{(\lambda E - p)}{cD(E)}\left[\left(\frac{2\lambda E + p}{c}\right) + 2E^2\left(\frac{\lambda E - p}{c^3}\right)\right] \tag{C.5}$$

and,

$$\frac{J(J+1)}{K(K+1)} - c.c. = \frac{i(\lambda E - p)}{cD(E)}\left[\left(1 - \frac{2E^2}{c^2}\right) + 3\lambda E\left(\frac{\lambda E - p}{c^2}\right)\right] \tag{C.6}$$

These relationships can be used in (3.25) and (3.27) along with (3.15) to obtain equations (3.30) and (3.31).